\documentclass[12pt]{article}
\usepackage{epsfig}
\usepackage{psfrag}
\usepackage{latexsym}
\def\be{\begin{equation}}
\def\ee{\end{equation}}
\def\bc{\begin{center}}
\def\ec{\end{center}}
\def\bea{\begin{eqnarray}}
\def\eea{\end{eqnarray}}
\def\dd{\displaystyle}
\def\nn{\nonumber}

\catcode`@=11
\def\marginnote#1{}
\newcount\hour
\newcount\minute
\newtoks\amorpm
\hour=\time\divide\hour by60
\minute=\time{\multiply\hour by60 \global\advance\minute by-\hour}
\edef\standardtime{{\ifnum\hour<12 \global\amorpm={am}%
        \else\global\amorpm={pm}\advance\hour by-12 \fi
        \ifnum\hour=0 \hour=12 \fi
        \number\hour:\ifnum\minute<10 0\fi\number\minute\the\amorpm}}
\edef\militarytime{\number\hour:\ifnum\minute<10 0\fi\number\minute}
\def\draftlabel#1{{\@bsphack\if@filesw {\let\thepage\relax
   \xdef\@gtempa{\write\@auxout{\string
      \newlabel{#1}{{\@currentlabel}{\thepage}}}}}\@gtempa
   \if@nobreak \ifvmode\nobreak\fi\fi\fi\@esphack}
        \gdef\@eqnlabel{#1}}
\def\@eqnlabel{}
\def\@vacuum{}
\def\draftmarginnote#1{\marginpar{\raggedright\scriptsize\tt#1}}
\def\draft{\oddsidemargin 0.0truein
        \def\@oddfoot{\sl preliminary draft \hfil
        \rm\thepage\hfil\sl\today\quad\militarytime}
        \let\@evenfoot\@oddfoot \overfullrule 3pt
        \let\label=\draftlabel
        \let\marginnote=\draftmarginnote
   \def\@eqnnum{(\theequation)\rlap{\kern\marginparsep\tt\@eqnlabel}%
\global\let\@eqnlabel\@vacuum}  }
\catcode`@=12
\begin{document}
\begin{titlepage}
\vspace*{-1cm}
\phantom{hep-ph/0207014} 
\hfill{DFPD-02/TH/16}
\vskip 2.0cm
\begin{center}
{\Large\bf Symmetry Breaking\\
\vskip .1cm
for Bosonic Systems on Orbifolds}
\end{center}
\vskip 1.5  cm
\begin{center}
{\large Carla Biggio}~\footnote{e-mail address: biggio@pd.infn.it}
\\
\vskip .1cm
Dipartimento di Fisica `G.~Galilei', Universit\`a di Padova and
\\ 
INFN, Sezione di Padova, Via Marzolo~8, I-35131 Padua, Italy
\\
\vskip .2cm
{\large Ferruccio Feruglio}~\footnote{e-mail address: feruglio@pd.infn.it}
\\
\vskip .1cm
Dipartimento di Fisica `G.~Galilei', Universit\`a di Padova and
\\ 
INFN, Sezione di Padova, Via Marzolo~8, I-35131 Padua, Italy
\end{center}
\vskip 1.5cm
\begin{abstract}
\noindent
We discuss a general class of boundary conditions for bosons living
in an extra spatial dimension compactified on $S^1/Z_2$.
Discontinuities for both fields and their first derivatives are 
allowed at the orbifold fixed points. We analyze examples
with free scalar fields and interacting gauge vector bosons,
deriving the mass spectrum, that depends on a combination of
the twist and the jumps. We discuss how the same physical system 
can be characterized by different boundary conditions, related by 
local field redefinitions that turn a twist into a jump or vice-versa. 
When the description is in term of discontinuous fields, appropriate 
lagrangian terms should be localized at the orbifold fixed points.
\end{abstract}
\end{titlepage}
\setcounter{footnote}{0}
\vskip2truecm

%
%

%
\bc
{\bf 1. Introduction}
\ec
Field theoretical models defined in more than 4 space-time dimensions
have recently received a considerable attention. They often occur
in suitable limits of more fundamental theories, such as string theories,
that naturally requires extra spatial dimensions, or in the so-called
``deconstruction'' \cite{deco} when a four dimensional field theory 
simulates an extra dimensional behaviour. In the presence of compact
extra dimensions a special role is played by orbifolds \cite{orbi}, 
since they provide
a simple theoretical framework to describe, at low energies, 
four-dimensional chiral fermions. The great interest in compact extra 
dimensions arises also from specific mechanisms of
symmetry breaking, which have no counterpart in four dimensional (4D)
theories. Symmetry breaking from Wilson lines \cite{hoso}, from non trivial
boundary conditions \cite{ss} or from orbifold projection may find relevant
applications when discussing the breaking of the electroweak symmetry 
\cite{ew,ewsusy}, of supersymmetry \cite{ewsusy,susyb1,class,susyb2} 
and of grand unified symmetries \cite{gut,kawa,gut1,gut2,gut3}.
              
In this paper we will discuss the spectrum of 5D bosonic theories
with the extra dimension compactified on the orbifold $S^1/Z_2$.
We will adopt a general class of boundary conditions on the
fields and their first derivatives, and we will analyze the 
corresponding mass spectrum. When dealing with the orbifold
$S^1/Z_2$, the boundary conditions are fully specified not only by
the periodicity of the field variables, as in the case of the circle
$S^1$, but also by the possible jumps of the fields across the 
orbifold fixed points. These jumps are forbidden on 
manifolds, where the fields are required to be smooth everywhere,
but are possible on orbifolds at the singular points, 
provided the physical properties of the system remain well
defined. 
This possibility, which has been recently studied
for fermions in 5 dimensions \cite{bfz1}, will be extensively discussed here
in section 2. We will proceed in sections 3, 4 and 5 to
describe in detail several examples, involving scalar fields and
gauge vector bosons. In all these examples
the mass spectrum depends on a combination 
of the twist (defining the field periodicity) and the jumps
occurring at the two orbifold fixed points.

In the examples discussed here, the discontinuities
of the field variables, allowed by the orbifold construction, have
no intrinsic physical meaning. Indeed, they can always be removed 
by means of a local field redefinition (possibly combined with a discrete
translation), that does not change the physics and leads to an
equivalent description in terms of smooth field variables.
These smooth fields are characterized by a new twist that
embodies the overall effect of the original twist and  
jumps \cite{bfz1}. 
 
Discontinuous fields are a natural ingredient of many orbifold
compactifications that have been discussed in the literature.
Indeed, they are strictly related to lagrangian
terms that are localized at the orbifold fixed points.
Whenever a quadratic term is localized at a point in the
extra dimension, the equations of motion, integrated in a small region
across that point, lead to discontinuous fields.
On the other hand, localized lagrangian terms are 
an important aspect of orbifold constructions. Even when the starting
theory has no such terms, they are generated by the renormalization
procedure via loop corrections \cite{loops}. 
They are relevant in the discussion of
the gauge anomalies of orbifold theories \cite{anomalies}. 
They also occur in many
phenomenological constructions 
\cite{ew,ewsusy,susyb1,class,susyb2,gut,kawa,gut1,gut2,gut3}. 
In the last section of this paper
we discuss, for bosonic systems, the form of these localized terms
that arise from our generalized boundary conditions. We will show
that, despite their highly singular behaviour, after appropriate
regularization, they are needed for the consistency of the theory. 

In summary, we analyze the properties of a class of bosonic systems
with an extra dimension compactified on $S^1/Z_2$. These systems can be 
equally described by smooth or by discontinuous fields. 
The physics is independent from the choice of variables. 
When discontinuous fields are used, the lagrangian contains terms 
localized at the orbifold fixed points and the spectrum depends on 
parameters that specify both the twist and the jumps of the fields. 
When smooth fields are adopted, there is no `brane' action and the 
spectrum depends on a new overall twist.   

%
%

\vspace{0.5cm}
\bc
{\bf 2. Boundary conditions for bosonic fields on $S^1/Z_2$}
\ec
We consider a generic 5-dimensional (5D) theory 
compactified on the orbifold $S^1/Z_2$,
with space-time coordinates $x^M\equiv (x^m,y)$ and metric
$\eta^{MN}=diag(-1,+1,+1,+1,+1)$. We can represent
the orbifold on the whole real axis, provided we identify points
related by a translation $T$ and a reflection $Z_2$, which, in a
suitable coordinate system, are given by:
\bea
T: & y \to y+ 2 \pi R\nn\\
Z_2: & y \to -y~~~,
\label{tz2}
\eea
where $R$ is the radius of $S^1$.
We introduce a set of $n$ real 5D bosonic fields $\Phi(x,y)$, which 
we classify in representations of the 4D Lorentz group. 
We define the $(T,Z_2)$ transformations of the fields by
\bea
\Phi(y+ 2 \pi R) &=& U_\beta~ \Phi(y)\nn\\
\Phi(-y) &=& Z~ \Phi (y)~~~~~ ,
\label{z2rep}
\eea
where $U_\beta$ and $Z$ are constant orthogonal matrices and 
$Z$ has the property $Z^2=1$.  It is not restrictive
for us to take a basis in which $Z$ is diagonal, 
$Z = diag \, (1,\ldots,1,-1,\ldots,-1)$.

We look for boundary conditions on the fields $\Phi(x,y)$ 
and their derivatives, in the general class 
\be
\left(
\begin{array}{c}
\Phi\\
\partial_y\Phi
\end{array}
\right)(\gamma^+) =
V_\gamma~\left(
\begin{array}{c}
\Phi\\
\partial_y\Phi
\end{array}
\right)(\gamma^-)~~~,
\label{gbc}
\ee
where $\gamma=(0, \pi,\beta)$, $\gamma^\pm=(0^\pm,\pi R^\pm, y^\pm)$, 
and $V_\gamma$ are constant $2n$ by $2n$ matrices. 
We have defined $0^\pm\equiv\pm\xi$, $\pi R^\pm\equiv\pi R\pm\xi$, 
$y^-\equiv y_0$ and $y^+\equiv y_0+2 \pi R$. Here $\xi$ is a small
positive parameter and $y_0$ is a generic point of the $y$-axis,
for convenience chosen between $-\pi R+\xi$ and $-\xi$.
We observe that eq. (\ref{z2rep}) implies a
specific form for the matrix $V_\beta$ in (\ref{gbc}).
For the time being we keep a generic expression for 
$V_\beta$, as well as for $V_{0,\pi}$.

We will now constrain the matrices $V_\gamma$,
by imposing certain consistency requirements on our theory.
The spectrum of the theory is determined by the eigenmodes
of the differential operator $-\partial^2_y$. In order to deal with
a good quantum mechanical system, we first demand that
the operator $-\partial^2_y$ is self-adjoint with respect to the 
scalar product:
\be
\langle\Psi \vert \Phi\rangle = 
\int_{y^-}^{0^-} d y~ \Psi^T(y) \Phi(y)+
\int_{0^+}^{\pi R^-} d y~ \Psi^T(y) \Phi(y)+
\int_{\pi R^+}^{y^+} d y~ \Psi^T(y) \Phi(y)~~~~~,
\label{sprod}
\ee
where the limit $\xi\to 0$
is understood. If we consider the matrix element 
$\langle\Psi \vert \left(-\partial^2_y\Phi\right)\rangle$
and we perform a partial integration we obtain:
\bea
\langle\Psi \vert \left(-\partial^2_y\Phi\right)\rangle &=& 
\langle \left(-\partial^2_y\Psi\right) \vert \Phi\rangle+\nn\\
&+ &
\left[\Psi^T(y)\partial_y \Phi(y)-\partial_y \Psi^T(y) \Phi(y)
\right]^{0^+}_{0^-}+~~~\nn\\
&+ &
\left[\Psi^T(y)\partial_y \Phi(y)-\partial_y \Psi^T(y) \Phi(y)
\right]^{\pi R^+}_{\pi R^-}+~~~\nn\\
&+ &
\left[\Psi^T(y)\partial_y \Phi(y)-\partial_y \Psi^T(y) \Phi(y)
\right]^{y^-}_{y^+}~~~~~,
\eea
where $[f(y)]^a_b=f(a)-f(b)$. A necessary condition for the self-adjointness
of the operator $-\partial^2_y$ is that 
the three square brackets vanish. However this is not sufficient,
in general. To guarantee self-adjointness, the
domain of the operator $-\partial^2_y$ should coincide with the domain
of its adjoint. In other words, we should impose
conditions on $\Phi(y)$ and its first derivative in such a way
that the vanishing of the unwanted contributions implies precisely
the same conditions on $\Psi(y)$ and its first derivative.

It is easy to show that, in the class of boundary conditions (\ref{gbc}),
this happens when
\be
V_\gamma J~ V_\gamma^T = J~~~~~
~~~~~~ \gamma\in (0, \pi , \beta)~~~,
\label{ug1}
\ee
where $J\equiv i \sigma^2$ is the symplectic form in the space 
$(\Phi,\partial_y\Phi)$. 
Eq. (\ref{ug1}) restricts $V_\gamma$ in the symplectic group
$Sp(2 n)$. The simplest possibility is offered by $V_\gamma=1$, 
for $\gamma=(0, \pi , \beta)$. In this case $U_\beta=1$, the fields
are periodic and continuous across the orbifold fixed points.
When $V_\beta\ne 1$, the field variables are not periodic and we have 
a {\em twist}. Such boundary conditions are characteristic of
the conventional Scherk-Schwarz mechanism \cite{ss}. 
When $V_0$ or $V_\pi $ differs from unity, the fields or their 
derivatives are not continuous across the fixed points and we have 
{\em jumps}. Therefore, in close analogy 
with the fermionic case, we find that the boundary conditions for bosons
allow for both twist and jumps \cite{bfz1}. 
At variance with the fermionic case, twist and jumps can also affect 
the first derivative of the field variables.
Moreover, the self-adjointness alone does not forbid boundary conditions
that mix the fields and their $y$-derivatives.
For instance, if we have a single real scalar field $\varphi(y)$,
and we parametrize the generic 2 by 2 matrix $V_\gamma$ as:
\be
V_\gamma=
\left(
\begin{array}{cc}
A_\gamma& B_\gamma\\
C_\gamma& D_\gamma
\end{array}
\right)~~~,
\ee
the condition (\ref{ug1}) reduces to $\det V_\gamma=1$, 
as expected since $Sp(2)$ and $SL(2,R)$ are isomorphic. If $B_\gamma$ and 
$C_\gamma$ are not both vanishing, 
the boundary conditions will mix $\varphi$ and $\partial_y \varphi$.

While the field variables and their first derivative may have
twist and jumps, we should require that physical quantities remain periodic
and continuous across the orbifold fixed points. This poses 
a further restriction on the matrices $V_\gamma$. If the theory
is invariant under global transformations of a group G,
we can satisfy this requirement by asking that the matrices
$V_\gamma$ are in a $2n$-dimensional representation of G. 
The choice of scalar product made in (\ref{sprod}) does not allow
to consider symmetry transformations of the 5D theory that mix fields and 
$y$-derivatives. 
Moreover, it is not restrictive to consider orthogonal 
representations of G on the space of real fields $\Phi$.
In this case, the solution to eq. (\ref{ug1}) reads
\be
V_\gamma=
\left(
\begin{array}{cc}
U_\gamma& 0\\
0& U_\gamma
\end{array}
\right)~~~
~~~~~~~~~~~~~U_\gamma\in {\rm G}~~~,
\label{ug2}
\ee
where $U_\gamma$ is in an orthogonal $n$-dimensional
representation of G.

Finally, we should take into account consistency conditions
among the twist, the jumps and the orbifold
projection. If we combine a translation $T$ with a reflection $Z_2$,
from the definition (\ref{tz2}) it immediately follows that
$T~Z_2~T=Z_2$ and we ask that this relation is faithfully 
represented by the operators $U_\beta$ and $Z$ \cite{class,fkpz,ssft}.
An analogous relation is also obtained if we combine a jump
with a reflection \cite{bfz1}. Finally, each of the two possible
jumps should commute with the translation $T$. We thus have:
\bea
\label{cond2}
U_\gamma~ Z~ U_\gamma &=& Z~~~
~~~~~~~~~~~~~~\gamma\in (0, \pi , \beta)~~~,\nn\\
\left[U_0,U_\beta\right]&=& 0~~~,\nn\\
\left[U_\pi,U_\beta\right]&=& 0~~~.
\label{ug3}
\eea
If $[Z,U_\gamma]=0$, then $U_\gamma^2=1$ and twist and/or jumps 
have eigenvalues $\pm 1$. In particular, if also $U_0$ and $U_\pi$ commute, 
there is a basis where they are 
all diagonal with elements $\pm 1$. In this special case
the boundary conditions do not involve any continuous parameter.
However $U_0$ and $U_\pi$ should not necessarily commute.
We will discuss an example in section 4. When $[Z,U_\gamma]\ne 0$ or when
$[U_0,U_\pi]\ne 0$, continuous parameters can appear in $U_\gamma$.

In summary, the allowed boundary conditions on $\Phi(y)$ are
specified in eq. (\ref{gbc}), with the matrices $V_\gamma$
satisfying the requirements (\ref{ug2},\ref{ug3}).

%
%

\vspace{0.5cm}
\bc
{\bf 3. One scalar field}
\ec
It is instructive to analyze in detail the case of a single massless
scalar field $\varphi(x,y)$, of definite parity, $Z=+ 1$ to begin with.  
We start by writing the equations of motion for $\varphi$ 
\be
-\partial^2_y \varphi = m^2 \varphi~~~,
\ee
in each region $y_q<y<y_{q+1}$ of the real line, where $y_q\equiv q \pi R$ and
$q\in Z$. We have defined the mass $m$ through the 4D equation 
$\partial^2\varphi= m^2 \varphi$. 
The solutions of these equations can be  glued by exploiting the boundary 
conditions $V_0$ and $V_\pi$, imposed at 
$y=y_{2q}$ and $y=y_{2q+1}$, respectively. Finally,
the spectrum and the eigenfunctions are obtained by requiring that
the solutions have the twist described by $V_\beta$.

The equation of motion remains invariant if we multiply $\varphi$
by $\pm 1$, so that the group G of global symmetries is a parity
(to be distinguished from the orbifold symmetry $Z_2$ that acts also on 
the $y$ coordinate). We have $V_\gamma=\pm1$. For instance, we are allowed
to consider either periodic or antiperiodic fields.
We start by analyzing the case of no jumps, $V_0=V_\pi=+1$.
The solutions of the equations of motion, up to an arbitrary
$x$-dependent factor, are
\be
\varphi_1(y)=\cos m y~~~~~~~~m~R=
\left\{
\begin{array}{ccc}
n& &V_\beta=+1\\
n+\dd\frac{1}{2}& &V_\beta=-1
\end{array}
\right.~~~,
\label{cos}
\ee
and $n$ is a non-negative integer.
It is interesting to compare the result for $V_\beta=-1$ 
with that obtained
by assuming a jump in $y=0$: $(V_0,V_\pi,V_\beta)=(-1,+1,+1)$.
We find
\be
\varphi_2(y)=\epsilon(y/2)~ \sin m y~~~~~~~m~R=
n+\frac{1}{2}~~~.
\label{epsin}
\ee
where $\epsilon(y)$ is the sign function on $S^1$.


\vspace{0.1cm}
\begin{table}[!ht]
\caption{Eigenvalues and eigenfunctions of $-\partial^2_y$, for
even fields $\varphi_k$ $(k=1,2,3,4)$ and odd fields $\varphi_k$ 
$(k=5,6,7,8)$. The spectrum is given in terms of the non-negative
integer $n$.
\label{tab1}}
\vspace{0.4cm}
\begin{center}
\begin{tabular}{|c|c|c|c|c|}   
\hline
& & & &\\                         
$(V_0,V_\pi,V_\beta)$ & $m~ R$ & {\tt eigenfunctions} & $\varphi(0)$ &
$\varphi(\pi R)$\\ 
& & & &\\
\hline
& & & &\\
$(+,+,+)$ &  $n\ge 0$ & $\varphi_1(y)=\cos m y$ & $\ne 0$ & $\ne 0$\\
 &  $n> 0$ & $\varphi_5(y)=\sin m y$ & $0$ & $0$\\
& & & &\\
\hline
& & & &\\
$(-,+,-)$ &  $n> 0$ & $\varphi_2(y)=\epsilon(y/2)~ \sin m y$ & 
$0$ & $0$\\
&  $n\ge 0$ & $\varphi_6(y)=\epsilon(y/2)~ \cos m y$ & 
${\tt jump}$ & $\ne 0$\\
& & & &\\
\hline
& & & &\\
$(+,-,-)$ &  $n\ge 0$ & $\varphi_3(y)=\epsilon(y/2+\pi R/2)\cos m y$ & 
$\ne 0$ & {\tt jump}\\
&  $n> 0$ & $\varphi_7(y)=\epsilon(y/2+\pi R/2)\sin m y$ & 
$0$ & 0\\
& & & &\\
\hline
& & & &\\
$(-,-,+)$ &  $n>0$ & $\varphi_4(y)=\epsilon(y)~ \sin m y$ & 
$0$ & $0$\\
&  $n\ge 0$ & $\varphi_8(y)=\epsilon(y)~ \cos m y$ & 
{\tt jump} &{\tt jump} \\
& & & &\\
\hline
& & & &\\
$(+,+,-)$ &  $n+1/2$ & $\varphi_1(y)=\cos m y$ & $\ne 0$ & $0$\\
&  $n+1/2$ & $\varphi_5(y)=\sin m y$ & $0$ & $\ne 0$\\
& & & &\\
\hline
& & & &\\
$(-,+,+)$ &  $n+1/2 $ & $\varphi_2(y)=\epsilon(y/2)~ \sin m y$ & 
$0$ & $\ne 0$\\
&  $n+1/2 $ & $\varphi_6(y)=\epsilon(y/2)~ \cos m y$ & 
{\tt jump}& $0$\\
& & & &\\
\hline
& & & &\\
$(+,-,+)$ &  $n+1/2$ & $\varphi_3(y)=\epsilon(y/2+\pi R/2)\cos m y$ & 
$\ne 0$ & 0\\
&  $n+1/2$ & $\varphi_7(y)=\epsilon(y/2+\pi R/2)\sin m y$ & 
$0$ & {\tt jump}\\
& & & &\\
\hline
& & & &\\
$(-,-,-)$ &  $n+1/2$ & $\varphi_4(y)=\epsilon(y)~ \sin m y$ & 
$0$ & ${\tt jump}$\\
&  $n+1/2$ & $\varphi_8(y)=\epsilon(y)~ \cos m y$ & 
${\tt jump}$& 0\\
& & & &\\
\hline
\end{tabular}
\end{center}
\end{table}


The eigenfunctions (\ref{cos}) and (\ref{epsin})
for $m=1/(2R)$ are compared in the third row of fig. 1. 
We observe that $V_0=+1$ implies 
$\partial_y\varphi=0$  at $y=0$, that is Neumann boundary conditions.
Instead, if we take $V_0=-1$, the even field $\varphi$ should vanish
at $y=0$ as for a Dirichelet boundary condition, and this
produces a cusp at $y=0$. Despite this difference, the
two eigenfunctions are closely related.
If compared in the region $0<y<\pi R$, they look the same, up
to an exchange between the two walls at $y=0$ and $y=\pi R$.
They both vanish at one of the two walls and they have the
same non-vanishing value at the other wall, with the same
profile in between.
%
%
%
\begin{figure}[!h]
\centerline{
\psfig{figure=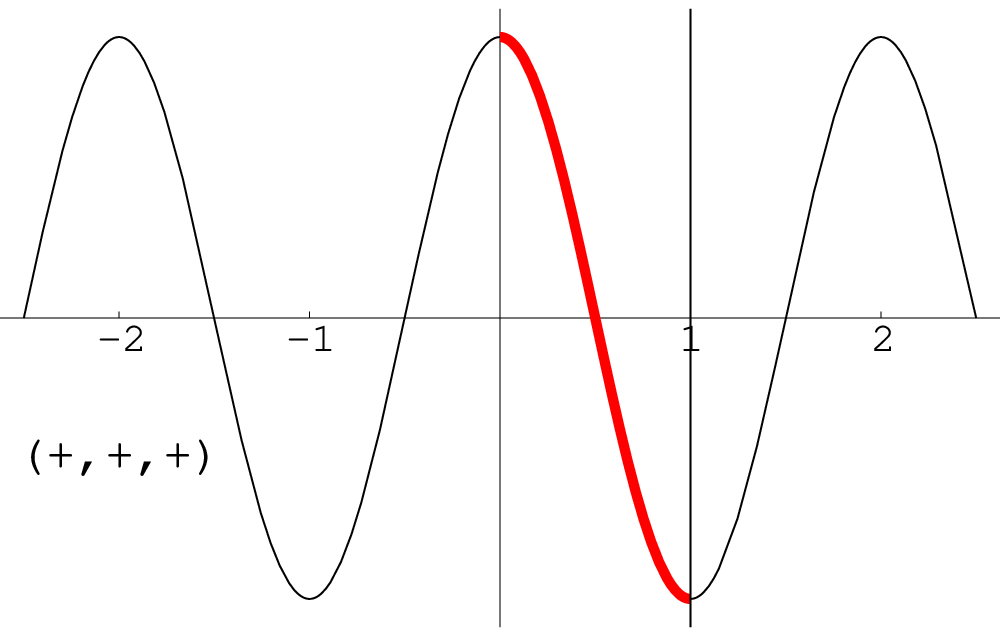,height=3cm,width=6cm}
\psfig{figure=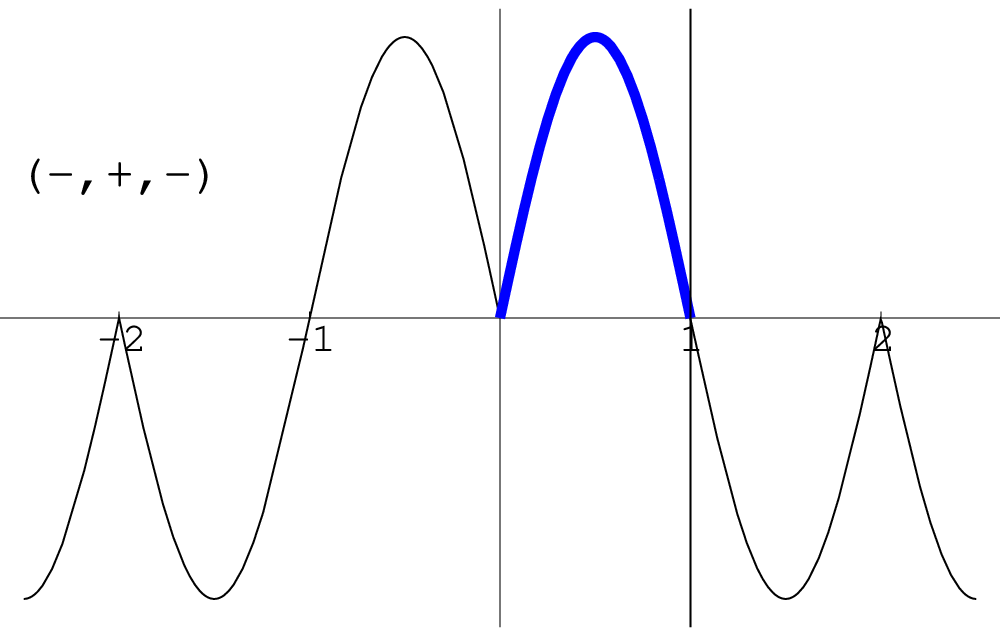,height=3cm,width=6cm}}
\vspace{0.3cm}
\centerline{
\psfig{figure=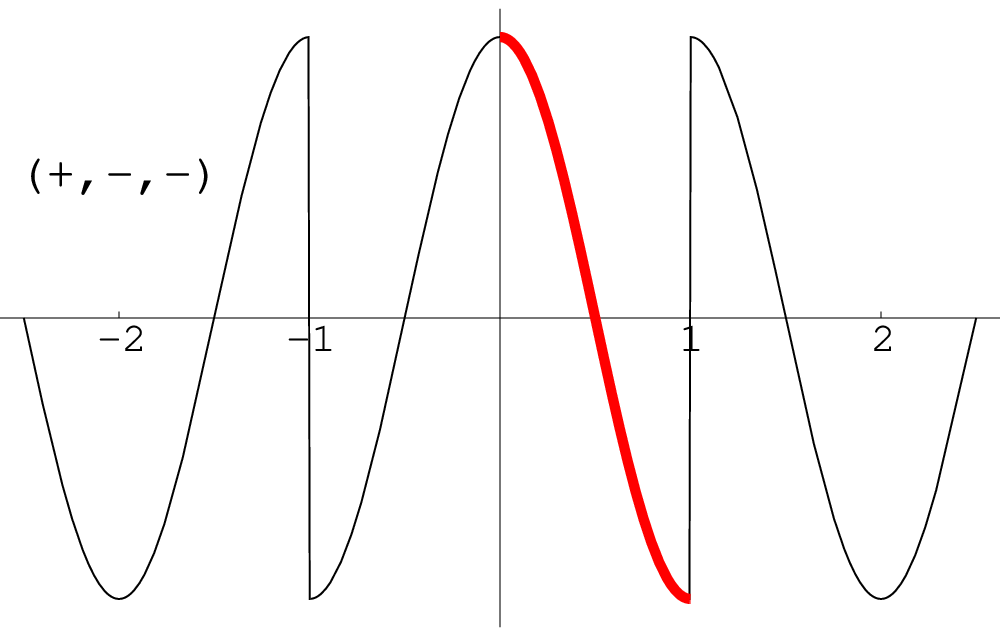,height=3cm,width=6cm}
\psfig{figure=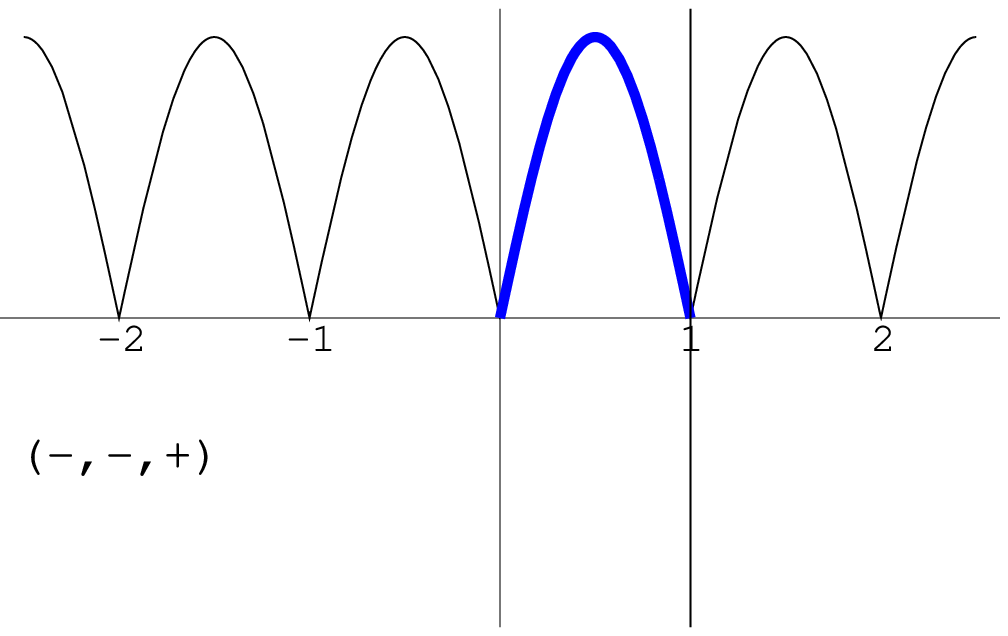,height=3cm,width=6cm}}
\vspace{0.3cm}
\centerline{
\psfig{figure=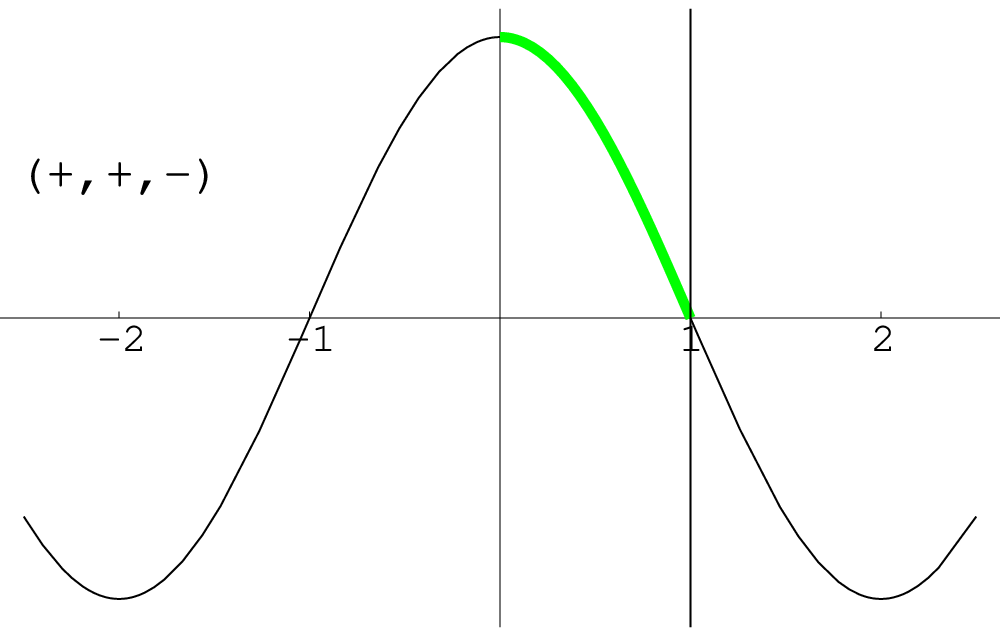,height=3cm,width=6cm}
\psfig{figure=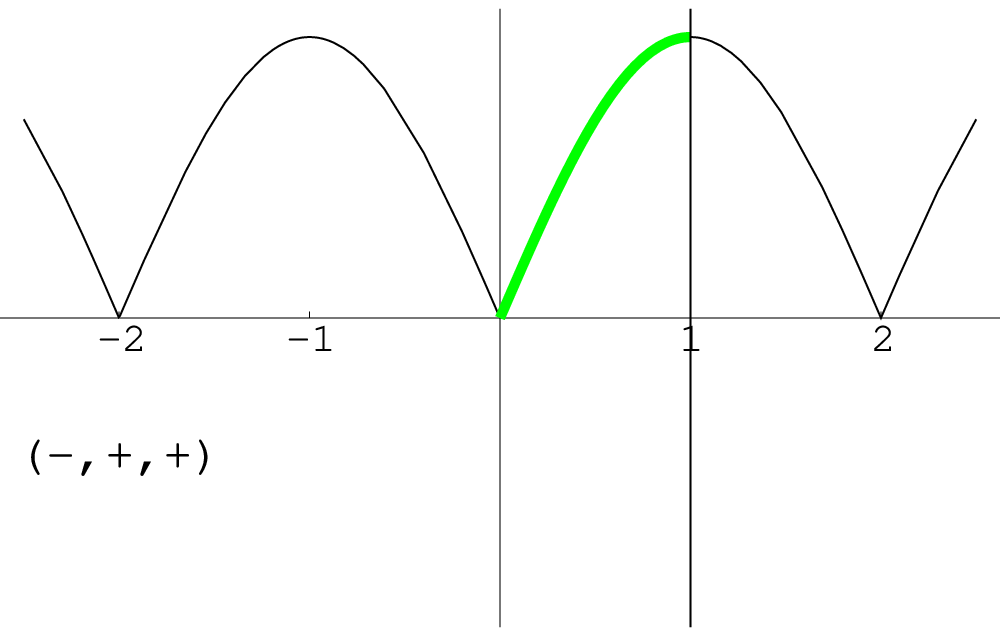,height=3cm,width=6cm}}
\vspace{0.3cm}
\centerline{
\psfig{figure=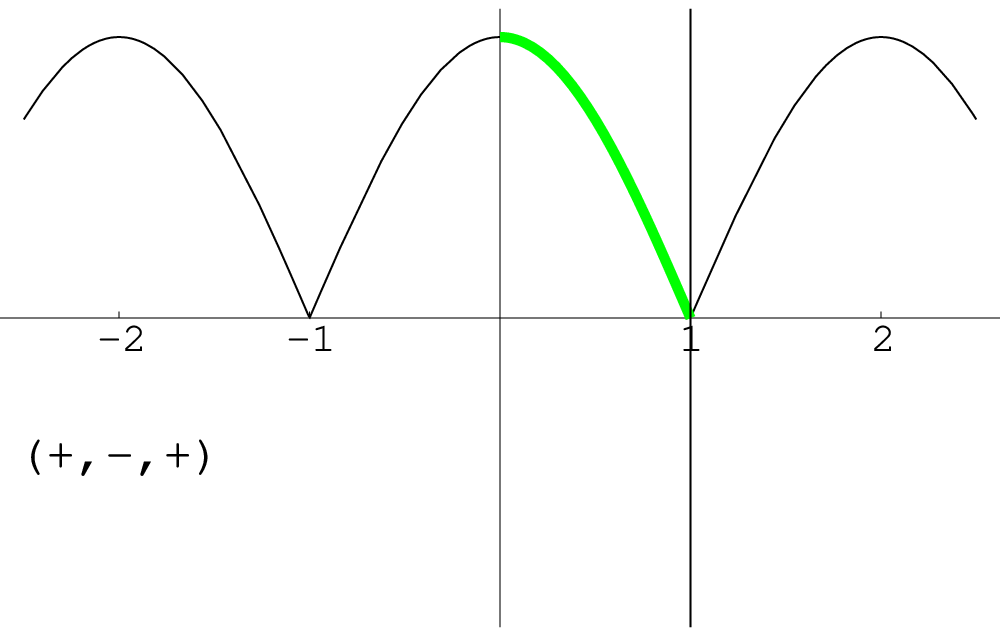,height=3cm,width=6cm}
\psfig{figure=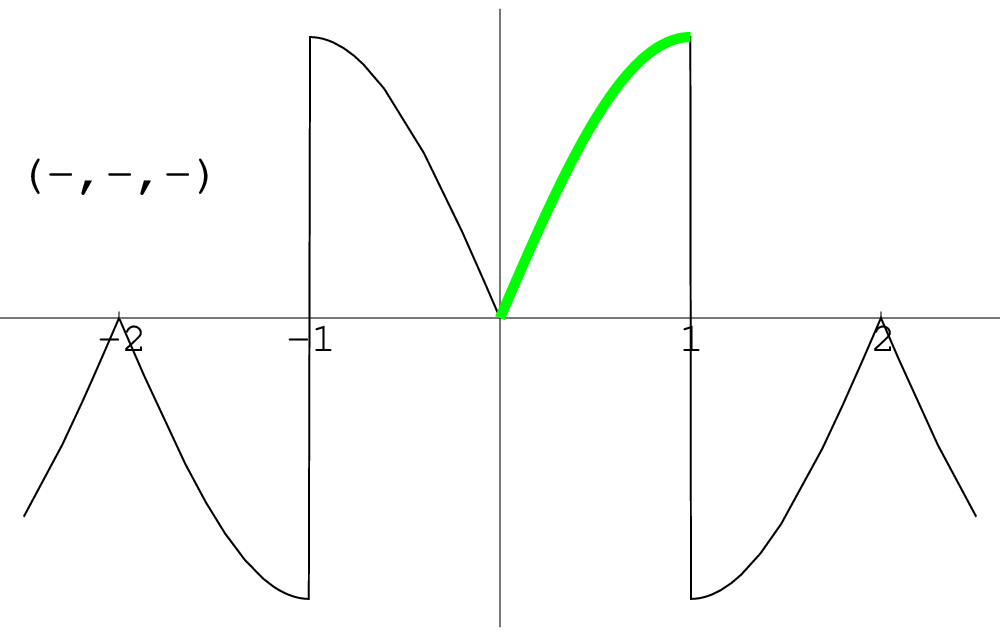,height=3cm,width=6cm}}
\caption{Eigenfunctions of $-\partial^2_y$, for a real even field $\varphi(y)$,
versus $y/(\pi R)$. For each boundary condition, labelled by 
($V_0,V_\pi,V_\beta$), the eigenfunction corresponding to the
lightest non-vanishing mode is displayed.
\label{evenf}}
\end{figure}
%
%
Indeed, the two cases are related by a coordinate transformation
and a field redefinition:
\be
\varphi_2(y)=\epsilon(y/2)~\varphi_1(y+\pi R)~~~.
\label{redef}
\ee
If $\varphi_1(y)$ is even, continuous and antiperiodic,
it is easy to see that the function $\varphi_2(y)$ defined
in (\ref{redef}) is even, periodic and has a cusp in $y=y_{2q}$,
where it vanishes. The equations of motion are not affected by the 
translation $y\to y+\pi R$, which simply exchanges the boundary 
conditions at $y=y_{2q}$ and $y=y_{2q+1}$. Moreover, the physical 
properties of a quantum mechanical system are invariant under
a local field redefinition. Therefore the two systems related
by eq. (\ref{redef}) are equivalent.

In table 1 we collect spectrum and eigenvalues for all possible
cases that are allowed by an even or odd field $\varphi$. 
We have found it useful to express the solutions in terms
of the sign function, which specifies the singularities
of the system. Indeed, $\epsilon(y/2)$ is singular in
$y=y_{2q}$, $\epsilon(y/2+\pi R/2)$ in $y=y_{2q+1}$ and 
$\epsilon(y)$ in all $y=y_q$. The correct parity of
the solutions is guaranteed by the properties of $\epsilon(y)$.
Also the periodicity can be easily determined from the
fact that $\epsilon(y)$ is periodic, whereas $\epsilon(y/2)$
and $\epsilon(y/2+\pi R/2)$ are antiperiodic.


\begin{figure}[!h]
\centerline{
\psfig{figure=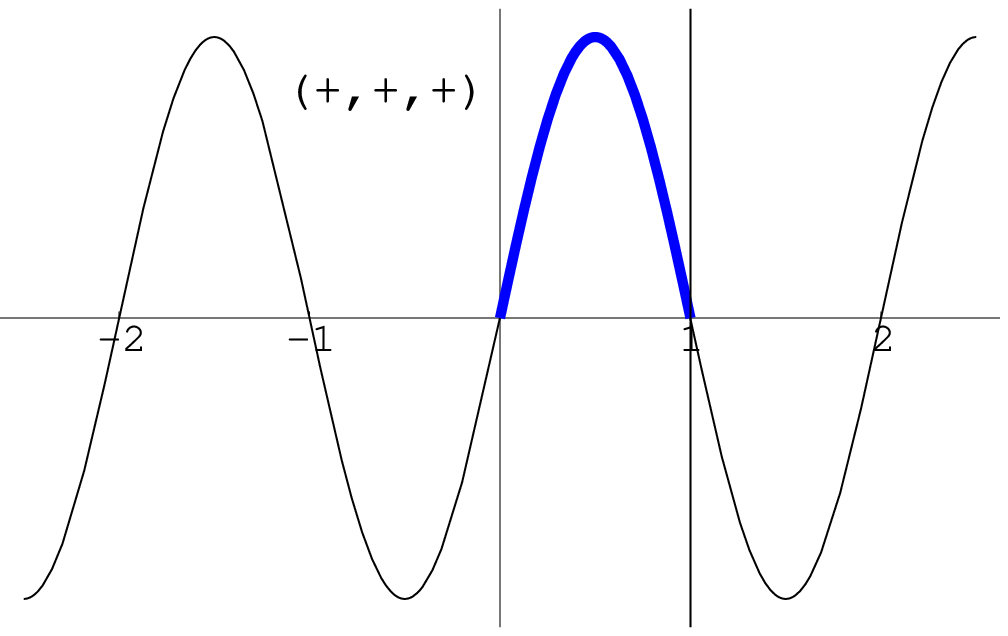,height=3cm,width=6cm}
\psfig{figure=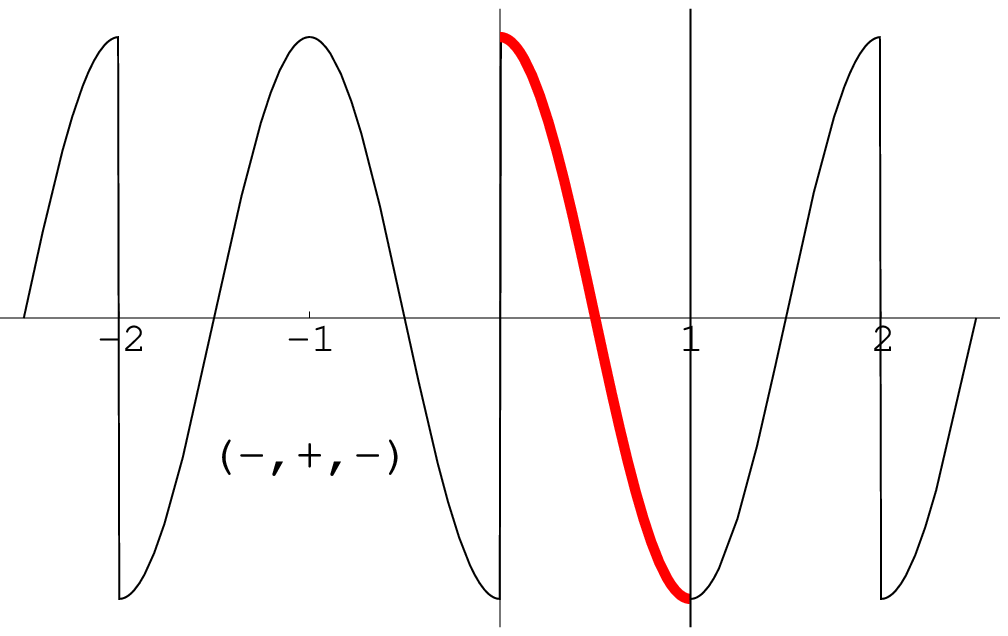,height=3cm,width=6cm}}
\vspace{0.3cm}
\centerline{
\psfig{figure=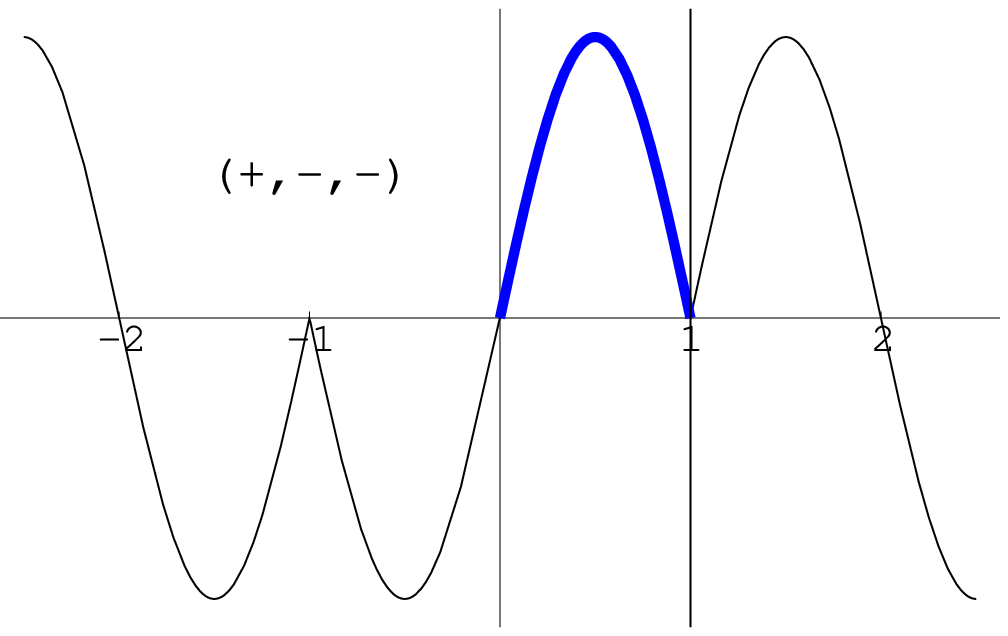,height=3cm,width=6cm}
\psfig{figure=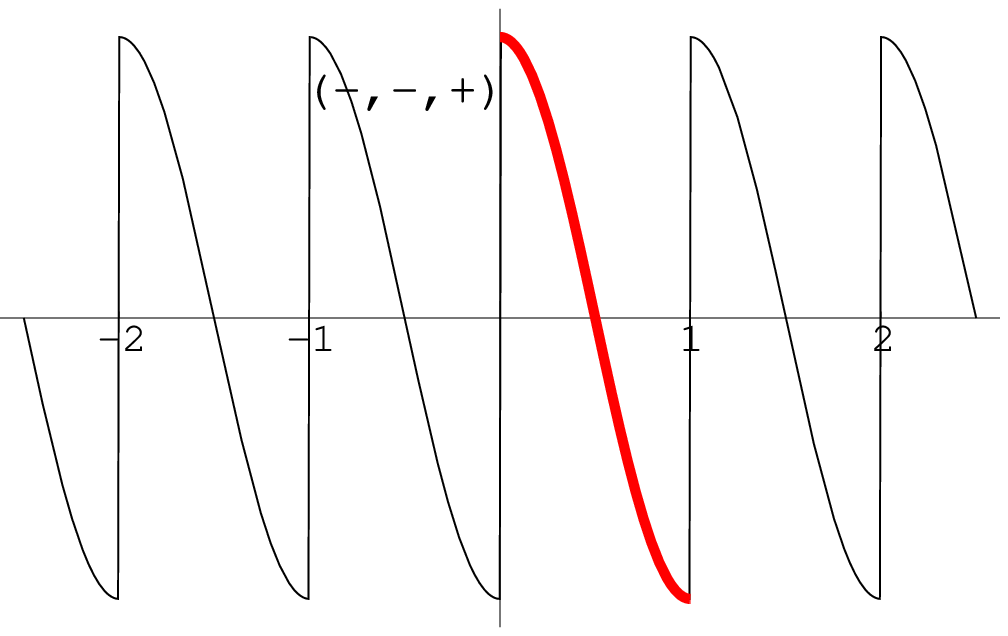,height=3cm,width=6cm}}
\vspace{0.3cm}
\centerline{
\psfig{figure=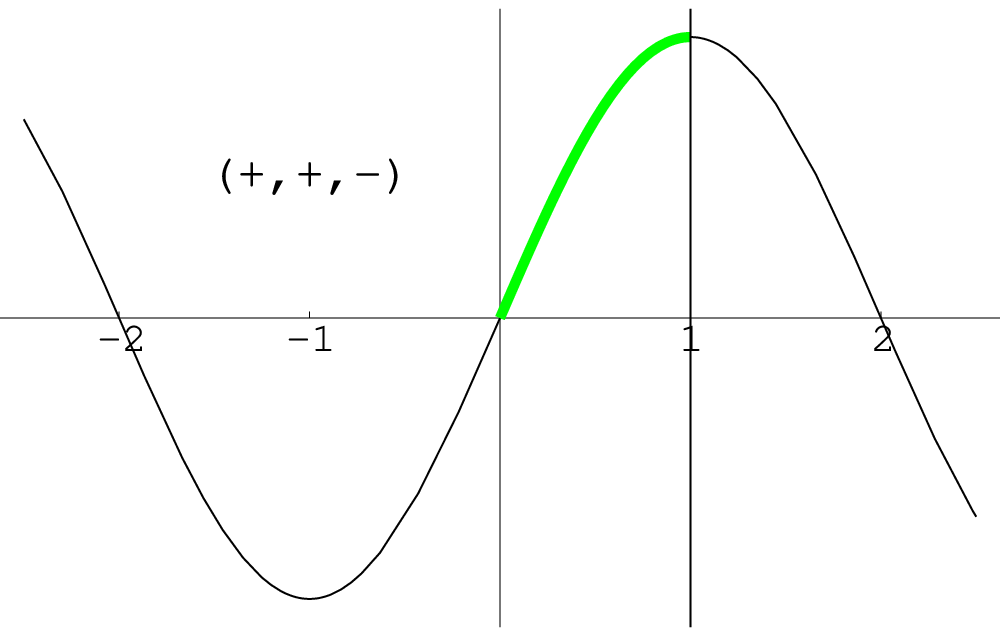,height=3cm,width=6cm}
\psfig{figure=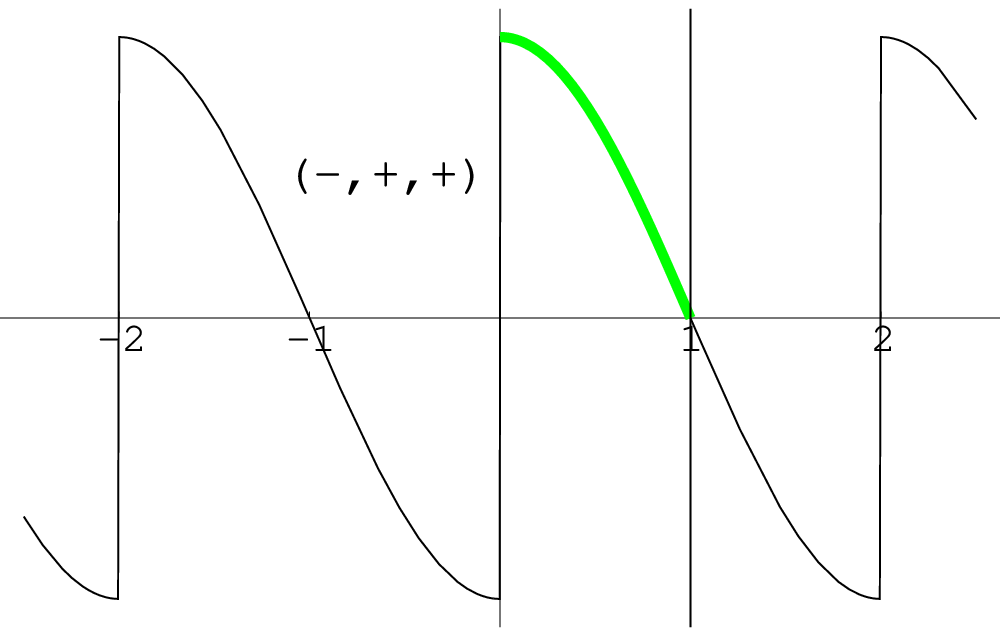,height=3cm,width=6cm}}
\vspace{0.3cm}
\centerline{
\psfig{figure=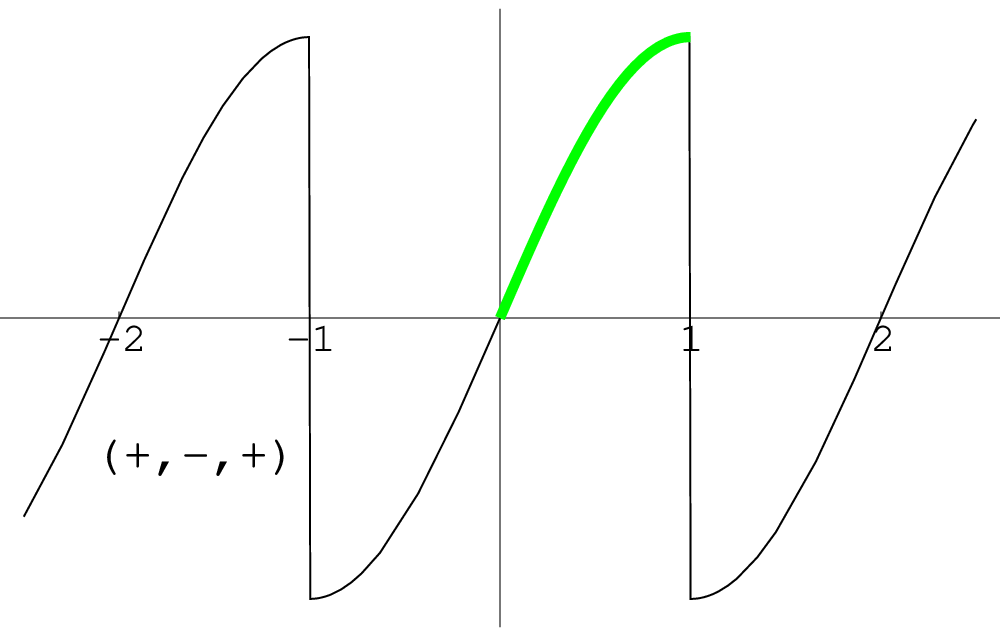,height=3cm,width=6cm}
\psfig{figure=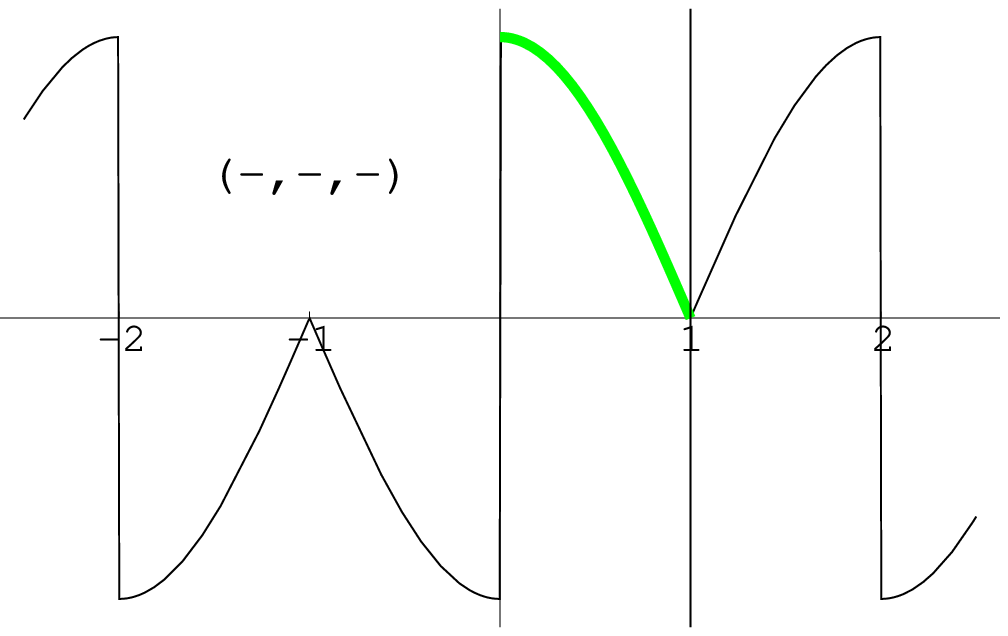,height=3cm,width=6cm}}
\caption{Eigenfunctions of $-\partial^2_y$, for a real odd field $\varphi(y)$,
versus $y/(\pi R)$. For each boundary condition, labelled by 
($V_0,V_\pi,V_\beta$), the eigenfunction corresponding to the
lightest non-vanishing mode is displayed.
\label{oddf}}
\end{figure}



There are three types of spectra: first, the ordinary Kaluza-Klein
tower $n/R$ that includes a zero mode; second, an identical spectrum
with the absence of the zero mode and, finally, the
Kaluza-Klein tower shifted by $1/2R$. All systems that
possess the same spectrum can be related by field redefinitions
that can be easily derived from table 1. The only non-trivial
transformation, applying only to the case of semi-integer
spectrum, is the one in (\ref{redef}). For semi-integer spectrum, 
all $\varphi_k$ $(k=1,...8)$ are related. For integer and non-negative
spectrum, we have maps among $\varphi_1$, $\varphi_3$, $\varphi_6$ and
$\varphi_8$. For strictly positive integer spectrum,
$\varphi_2$, $\varphi_4$, $\varphi_5$ and $\varphi_7$ are related.
Thanks to these relations, we can always go from a description in terms 
of discontinuous field variables to a descriptions by the smooth fields 
$\varphi_1(y)$ or $\varphi_5(y)$. 
Also, as can be seen from figs. 1 and 2, the behavior
in the vicinity of the fixed points is the same for all the eigenfunctions 
representing the same type of spectrum, up to a possible
exchange between the two fixed points.
In the presence of a single real field $\varphi$ the parity $Z$ 
does not seem to have an absolute physical meaning. We
find that there are equivalent physical systems with
opposite $Z$ parities for $\varphi$.

In conclusion, there are less physically inequivalent
systems than independent boundary conditions, at
least for free theories. There are different boundary
conditions that lead to the same spectra and the corresponding
systems are related by field redefinitions. The parameter
$V_0\cdot V_\pi\cdot V_\beta$ is equal in equivalent systems.
When $V_0\cdot V_\pi\cdot V_\beta=+1$, $m~R$ is integer
whereas for $V_0\cdot V_\pi\cdot V_\beta=-1$, $m~R$
is semi-integer. 

%
%

\vspace{0.5cm}
\bc
{\bf 4. More scalar fields}
\ec
Several scalar fields lead to the possibility of exploiting
continuous global symmetries to characterize boundary conditions.
As an example, we consider here the case of a 5D complex scalar
field $\varphi(y)\equiv (\varphi_1(y) + i~ \varphi_2(y))/\sqrt{2}$.
Its equation of motion:
\be
-\partial^2_y \varphi = m^2 \varphi~~~,
\ee
is invariant under global O(2) transformations, acting
on $(\varphi_1,\varphi_2)$.

We first discuss the case where $Z=diag(+1,-1)$ in the basis 
$(\varphi_1,\varphi_2)$. In this case we can take:
\be
U_\gamma=
\left(
\begin{array}{cc}
\cos\theta_\gamma & \sin\theta_\gamma\\
-\sin\theta_\gamma & \cos\theta_\gamma
\end{array}
\right)~~~,
\label{rotation}
\ee
which is a symmetry of the theory and satisfies (\ref{ug3}).
In general, we can choose three independent angles 
$\theta_\gamma=(\beta,\delta_0,\delta_\pi)$
for the twist and the two jumps at $y=0,\pi R$ respectively.
The solution of the equation of motion subjected to these
boundary conditions can be obtained by the same method
used in section 3. We find:
\be
\varphi(y)=e^{\dd i (m y - \alpha(y))}~~~,
\label{sol2}
\ee
where
\be
m=\frac{n}{R} - \frac{(\beta-\delta_0-\delta_\pi)}{2 \pi R}~~~
~~~~~~~(n\in Z)~~~,
\ee
and
\be
\alpha(y)=\frac{\delta_0-\delta_\pi}{4}~\epsilon(y)
+\frac{\delta_0+\delta_\pi}{4}~\eta(y)~~~.
\label{alpha}
\ee
The function $\eta(y)$ is the `staircase' function
\be
\eta(y)=2 q+1~~~~~~~~~y_q<y<y_{q+1}~~~~~~~~~~(q\in Z)~~~.
\label{staircase}
\ee
The function $\alpha(y)$ is flat everywhere but at $y=y_q$,
where it jumps, leading to the desired behavior of
the solution at the fixed points. It
satisfies $\alpha(y+2 \pi R)=\alpha(y)+\delta_0+\delta_\pi$,
thus contributing to the overall twist $\beta$ of $\varphi(y)$
by the amount $\delta_0+\delta_\pi$. This explains why the 
shift of the spectrum with respect to the Kaluza-Klein
levels is given by $\beta-\delta_0-\delta_\pi$ and not
by $\beta$ as in the conventional Scherk-Schwarz
mechanism. 

When $\beta-\delta_0-\delta_\pi=0~~~({\rm mod}~ 2\pi)$, the masses are $n/R$ 
and we can order all
massive modes in pairs. Indeed each physical non-vanishing mass
$\vert m\vert=\vert n\vert/R$ $(n\ne 0)$ corresponds to two independent eigenfunctions.
For instance, when $\beta=\delta_0=\delta_\pi=0$, we have
$\varphi_{\pm}^n=\exp(\pm i n y/R)$.
This infinite series of degenerate 4D doublets can be interpreted
as a consequence of the $O(2)$ symmetry, which is unbroken.
A non-vanishing shift of the Kaluza-Klein levels induces an explicit
breaking of the $O(2)$  symmetry. The order parameter
is $\beta-\delta_0-\delta_\pi$ $({\rm mod}~ 2\pi)$. 
When $\beta-\delta_0-\delta_\pi$
is non-vanishing (and not a multiple of $2 \pi$), 
the eigenfunctions of the massive modes are no longer
paired, each of them corresponding now to a different physical mass.
As for the case of a single
real field, different boundary conditions may lead
to the same spectrum. For instance, it is possible that $O(2)$
remains unbroken, despite the existence of non-trivial
boundary conditions, if twist and jumps are such 
that the combination $\beta-\delta_0-\delta_\pi$ 
vanishes mod $2\pi$. Moreover, starting from a generic system with
both twist $\beta$ and jumps $\delta_{0,\pi}$ different from 
zero, we can always move to an equivalent `smooth'
theory where the jumps vanish and the twist $\beta^c$
of the new scalar field $\varphi^c(y)$ is given by 
$\beta-\delta_0-\delta_\pi$. The map between the two systems 
is given by:
\be
\varphi^c(y)=e^{\dd{i\alpha(y)}}\varphi(y)~~~.
\label{map}
\ee
The multiplicative factor $e^{\dd{i\alpha(y)}}$ removes
the discontinuities from $\varphi(y)$ and add a twist $-\delta_0
-\delta_\pi$ to the wave function. 

Another interesting case is that of $Z$
proportional to the identity.
If we assign the same $Z$ parity to the real components 
$\varphi_{1,2}(y)$, then $U_\gamma$ commutes with $Z$
and the condition (\ref{ug3}) implies that its eigenvalues 
are $\pm 1$. If also $[U_0,U_\pi]=0$, then it is not
restrictive to go to a field basis where all $U_\gamma$
are diagonal, with elements $\pm 1$. 
This would lead to a discussion qualitatively 
close to that of section 3, where twist and jumps 
were quantized. A new feature occurs if
$[U_0,U_\pi]\ne 0$. Consider as an example $Z=U_\beta=diag(+1,+1)$
in the basis $(\varphi_1,\varphi_2)$. A consistent choice
for $U_0$ and $U_\pi$ is:
\be
U_0=
\left(
\begin{array}{cc}
\cos\delta_0 & -\sin\delta_0\\
-\sin\delta_0 & -\cos\delta_0
\end{array}
\right)~~~,~~~~~~~~~
U_\pi=
\left(
\begin{array}{cc}
1 & 0\\
0& -1
\end{array}
\right)~~~.
\label{ncjumps}
\ee 
Notice that the $O(2)$ matrices $U_0$ and $U_\pi$
square to 1, as required by the condition (\ref{ug3}).
The solutions of the equations of motion are:
\bea
\phi_1(y)&=&\cos(my-\alpha'(y))\nn\\
\phi_2(y)&=&\epsilon(y)\sin(my-\alpha'(y))~~~,
\label{sol3}
\eea
where
\be
m=\frac{n}{R} + \frac{\delta_0}{2 \pi R}~~~
~~~~~~~(n\in Z)~~~,
\ee
and
\be
\alpha'(y)=\frac{\delta_0}{4}~\left(\epsilon(y)
+\eta(y)\right)~~~.
\label{alphap}
\ee
It is interesting to note that this choice of boundary conditions
leads to a theory that is physically equivalent to 
that studied at the beginning of this section, where
the fields $\varphi_1$ and $\varphi_2$ had opposite parity.
We can go back to that system and consider the case 
of periodic fields with a jump at $y=0$:
$Z=diag(+1,-1)$, $U_\pi=U_\beta=1$ and $U_0$ as in (\ref{rotation}) 
with $\theta_0=\delta_0$. If we now perform the field redefinition:
\be
\varphi_1(y)\to\varphi_1(y)~~~,~~~~~~~~~~~
\varphi_2(y)\to\epsilon(y)~\varphi_2(y)~~~,
\ee
the new field variables are both even and periodic
and their jumps are those given in (\ref{ncjumps}).
It is easy to see that also the solutions (\ref{sol2})
are mapped into (\ref{sol3}). Moreover, it will be now
possible to describe the theory defined by the jumps
(\ref{ncjumps}) in terms of smooth field variables,
characterized by a certain twist. 

This correspondence provides another example
of equivalent systems, despite a different assignment of the 
orbifold parity. 
The presence of discontinuous fields is a generic feature
of field theories on orbifolds. The present discussion suggests that
at least in some cases these discontinuities may not have 
any physical significance, being only related to a particular
and not compelling choice of field variables.   
\vfill
\newpage

%
%

\bc
{\bf 5. Gauge vector bosons}
\ec
Our generalized boundary conditions can be exploited to spontaneously break
the gauge invariance of a 5D system. This is well-known as far as the
twist is concerned. A non-trivial twist induces a shift in the 
Kaluza-Klein levels. This lifts the zero modes of the gauge vector
bosons and the gauge symmetry, from a 4D point of view, is broken 
\cite{hoso,ss}.
As we have seen in the previous sections, also the discontinuities
of the fields and their first derivatives have a similar effect
on the spectrum and we may expect that, in the context of a gauge
theory, they lead to spontaneous breaking of the 4D gauge invariance.
To analyze these aspects, we focus on a 5D gauge theory defined
on our orbifold and based on the gauge group SU(2).

Not all $Z_2$ parity assignments for the gauge fields $A^a_M(x,y)$
$(a=1,2,3)$, $(M=\mu,5)$, $(\mu=0,1,2,3)$ are now possible. The gauge invariance
imposes several restrictions. First of all, the action of $Z_2$
on the 4D vector bosons $A^a_\mu$ should be compatible with the
algebra of gauge group. In other words, we should embed $Z_2$
into the automorphism algebra of the gauge group \cite{gut2}.
For SU(2) this leaves two possibilities: either all $A^a_\mu$
are even, or two of them are odd and one is even. Furthermore,
a well-defined parity for the field strength implies that 
the parity of $A^a_5$ should be opposite to that of $A^a_\mu$.
In the basis $(A^1_\mu,A^2_\mu,A^3_\mu)$, up to a re-labelling of
the 3 gauge fields, we can consider:
\bea
({\rm A})~~~~~~~~~~~~~Z&=&diag(+1,+1,+1)\nn\\
({\rm B})~~~~~~~~~~~~~Z&=&diag(-1,-1,+1)~~~.
\label{parity}
\eea
The boundary conditions on the $A^a_\mu$ are specified
by 3 by 3 matrices $U_\gamma$ that satisfy the consistency relation (\ref{ug3})
and leave the SU(2) algebra invariant \cite{class,gut2}. 
This last requirement can be
fulfilled by requiring that $U_\gamma$ is an SU(2) global transformation
that acts on $(A^1_\mu,A^2_\mu,A^3_\mu)$ in the adjoint representation.
Finally, to preserve gauge invariance, the boundary conditions on the scalar 
fields $A^a_5$ should be the same as those on the corresponding
4D vector bosons. This can be seen by asking that the various 
components of the field strength $F^a_{MN}$ possess well-defined 
boundary conditions. 

For instance, in the case (A) where all fields $A^a_\mu$ 
have even $Z_2$ parity, a consistent assignment is \cite{kawa}:
\be
U_\beta = diag(-1,-1,+1)~~~,~~~~~~~~~~U_0=U_\pi=diag(+1,+1,+1)~~~.
\label{bcg1}
\ee
In the gauge $\partial^M A^a_M=0$, the 5D
equation of motion read:
\be
-\partial^2_y A^a_M = m^2 A^a_M~~~.
\label{eomg}
\ee
The solutions with the appropriate boundary conditions are:
\be
A^a_\mu(x,y)=A^{a(n_a)}_\mu(x) \cos m_a y
~~~~~~m_{1,2} R = n_{1,2}+\frac{1}{2}
~~~~~~m_3 R= n_3
~~~,
\label{spg1}
\ee
where $n_{1,2,3}$ are non-negative integers.
The only zero mode of the system is $A^{3(0)}_\mu(x)$ and, from a 
4D point of view the original gauge symmetry is broken down to
the U(1) associated to this massless gauge vector boson.
The breaking of the 5D SU(2) gauge symmetry is spontaneous
and each mode in (\ref{spg1}), but $A^{3(0)}_\mu(x)$, 
becomes massive via an Higgs mechanism.
The unphysical Goldstone bosons are the modes of the fields
$A^a_5(x,y)$, which are all absorbed by the corresponding
massive vector bosons. On the wall at $y=0$ all the gauge 
fields and the parameters of the gauge transformations
are non-vanishing. Here all the constraints coming from the full 
5D gauge invariance are effective. Instead, on the wall at $y=\pi R$,
only $A^3_\mu(x,y)$ and the corresponding gauge parameter are
different from zero. Therefore the effective symmetry
at the fixed point $y=\pi R$ is the U(1) related to the
4D gauge boson $A^3_\mu(x,\pi R)$. 
This kind of setup where the gauge symmetry is broken by
twisted orbifold boundary conditions and the two fixed points 
are characterized by two different effective 4D symmetries
has recently received lot of attention, for its successful application in the
context of grand unified theories \cite{kawa,gut1,gut2,gut3}. 

It is interesting to note that the same physical system discussed
above can be described by using periodic field variables,
with discontinuities at the fixed points. This is achieved, for instance,
by means of the boundary conditions
\be
U_\beta = U_0=diag(+1,+1,+1)~~~,~~~~~~~~~~U_\pi=diag(-1,-1,+1)~~~.
\label{bcg2}
\ee
The solutions to the equations of motion (\ref{eomg}) are:
\bea
A^a_\mu(x,y)&=&A^{a(n_a)}_\mu(x)~ \epsilon(y/2+\pi R/2)\cos m_a y
~~~~~~m_a R = n_a+\frac{1}{2}~~~~~~~~~~(a=1,2)\nn\\
A^3_\mu(x,y)&=&A^{3(n_3)}_\mu(x) \cos m_3 y~~~~~~~~~~~~~~~m_3 R= n_3~~~,
\label{spg2}
\eea
where $n_{1,2,3}$ are non-negative integers. The new 
solutions $A^{1,2}_\mu$ have cusps at $y=y_{2q+1}$, as the profiles
denoted by $(+,-,+)$ in fig. 1.
The two descriptions are related by the field redefinition:
\be
A^a_\mu(x,y)~\to~ \epsilon(y/2+\pi R/2)~ A^a_\mu(x,y)~~~~~~(a=1,2)~~~.
\ee

In the previous example the boundary conditions $U_\gamma$ commute among themselves
and with the parity $Z$. As a consequence the rank of the gauge group SU(2)
is conserved in the symmetry breaking. We can lower the rank by assuming 
$[U_\gamma,Z]\ne 0$ \cite{gut2,gut3}. As an example, we consider the parity (B) of eq. (\ref{parity})
and boundary conditions described by:
\be
U_\gamma=e^{\dd \theta_\gamma T^2}~~~~~~~
T^2=
\left(
\begin{array}{ccc}
0& 0& 1\\
0& 0& 0\\
-1& 0& 0
\end{array}
\right)~~~,
\label{bg3}
\ee
in the basis $(A^1_\mu,A^2_\mu,A^3_\mu)$. We allow, at the same time, for 
a twist $\theta_\beta\equiv\beta$ and two jumps $\theta_{0(\pi)}\equiv\delta_0(\delta_\pi)$.
The matrices $U_\gamma$ are block diagonal and do not mix the index 2 with the indices (1,3).
Thus the boundary conditions are trivial for the odd field $A^2_\mu$ and its derivative.
Non-trivial boundary conditions involve the fields $A^1_\mu$ and $A^3_\mu$.
By solving the equations (\ref{eomg}), we obtain:
\bea
A^1_\mu(x,y)&=&A^{(n)}_\mu(x)~\sin(m y-\alpha(y))\nn\\ 
A^2_\mu(x,y)&=&A^{2(n_2)}_\mu(x)~\sin m_2 y~~~~~~~~~~~~~~~~~~~~~~~m_2 R=n_2~~~~\nn\\
A^3_\mu(x,y)&=&A^{(n)}_\mu(x)~\cos(m y-\alpha(y))~~~~~~~~~~~~~~~~m R = n - 
\frac{\beta-\delta_0-\delta_\pi}{2 \pi}~~~,
\eea
where $n\in Z$, $n_2$ is a positive integer and the function $\alpha(y)$ has been defined in eqs.
(\ref{alpha}) and (\ref{staircase}). If $\beta-\delta_0-\delta_\pi=0$ 
(mod $2\pi$), we have a zero mode $A^{(0)}_\mu(x)$
and the gauge symmetry is spontaneously broken down to U(1), as in the previously discussed examples.
When $\beta-\delta_0-\delta_\pi\ne0$ (mod $2\pi$), there are no zero modes and SU(2) is completely broken. 
We can go continuously from this phase to the phase where a U(1) survives, by changing the twist and/or
the jump parameters. We may thus have a situation where U(1) is broken by a very small amount,
compared to the scale $1/R$ that characterizes the SU(2) breaking. The U(1) breaking order parameter
is the combination $\beta-\delta_0-\delta_\pi$. The same physical system
is described by a double infinity of boundary conditions, those that reproduce the same order
parameter. All these descriptions are equivalent and are related by field redefinitions. In the class
of all equivalent theories one of them is described by continuous fields $A^{a~c}_\mu(x,y)$.
We go from the generic theory described in terms of $(\beta,\delta_0,\delta_\pi)$ to that
characterized by $(\beta^c\equiv\beta-\delta_0-\delta_\pi,\delta^c_0\equiv 0,\delta^c_\pi\equiv 0)$, via the
field transformation:
\be
\left(
\begin{array}{c}
A^{1~c}_\mu\\
A^{3~c}_\mu
\end{array}
\right)
=
\left(
\begin{array}{cc}
\cos\alpha(y)&-\sin\alpha(y)\\
\sin\alpha(y)&\cos\alpha(y)\\
\end{array}
\right)
\left(
\begin{array}{c}
A^{1}_\mu\\
A^{3}_\mu
\end{array}
\right)~~~.
\ee
\vspace{0.5cm}

\bc
{\bf 6. Brane action for bosonic system}
\ec

In the previous sections we showed the equivalence between bosonic systems 
characterized by discontinuous fields and `smooth' systems in which fields 
are continuous but twisted. 
For each pair of systems characterized by the same mass spectrum
we were able to find a local field redefinition, plus a possible 
discrete translation, mapping the mass eigenfunctions
of one system into those of the other system.
Here we would like to further explore the relation between smooth
and discontinuous systems by showing that the field discontinuities 
are strictly related to lagrangian terms localized at the fixed points.


We begin by discussing the case of one real scalar field. 
To fix the ideas we focus on the equivalence between the cases
$(+,+,-)$ and $(+,-,+)$ with $Z=1$ of table 1. The other cases can
be discussed along similar lines. We denote by $\varphi^c$ the continuous 
field $(+,+,-)$ with twist $U_\beta=-1$ and by $\varphi$ the periodic field
$(+,-,+)$ that has a jump $U_\pi=-1$. If we start from
the Lagrangian $\mathcal{L}$ for the boson $\varphi^{c}$
\be
\mathcal{L}(\varphi^{c},\partial\varphi^{c})=-\frac{1}{2}
\partial_{M}\varphi^{c} \partial^{M}\varphi^{c}~~~,
\ee
and we perform the field redefinition:
\be
\varphi^{c}(y) = \chi(y) \varphi(y)~~~~~~~~
\chi(y)\equiv\epsilon(y/2+\pi R/2)~~~,
\ee 
we obtain an expression in terms of discontinuous fields and their
derivatives, from which it is difficult to derive the correct equation 
of motion for the system. Indeed the new lagrangian is highly
singular and the naive use of the variational principle, 
which is tailored on continuous functions and smooth functionals, 
would lead to inconsistent results. 
In order to avoid these problems we regularize $\chi(y)$
by means of a smooth function $\chi_\lambda(y)$ ($\lambda>0$) which 
reproduces $\chi(y)$ in the limit $\lambda\to 0$. 
By performing the substitution:
\be
\varphi^{c}(y) = \chi_{\lambda}(y)~\varphi(y)~~~,
\ee
we obtain 
\be
\label{regl}
\mathcal{L}(\varphi^{c},\partial\varphi^{c}) =
-\frac{1}{2}\chi_{\lambda}^{2} \partial_{M}\varphi \partial^{M}\varphi -
\chi_{\lambda}\chi_{\lambda}'\varphi\partial_{y}\varphi -
\frac{1}{2}\chi_{\lambda}'^{2} \varphi^{2}~~~. 
\ee
Since the field $\varphi(y)$ is periodic, we can work in the
interval $0\le y \le 2\pi R$. In the limit 
$\lambda\to 0$, we find:
\be
\mathcal{L}(\varphi^{c},\partial\varphi^{c}) =
-\frac{1}{2} \chi^{2} \partial_{M}\varphi \partial^{M}\varphi  
+ 2\chi~ \delta_{\pi R}~ \varphi\partial_{y}\varphi - 
 2 \delta_{\pi R}^{2}~ \varphi^{2}~~~~~~~~~y\in [0,2\pi]~~~,
\ee
with $\delta_{y_{0}} \equiv \delta (y-y_{0})$.

The action contains quadratic terms for the field $\varphi(y)$ that
are localized at $y=\pi R$. 
However these terms are quite singular
and, strictly speaking, are mathematically ill-defined even as distributions.
For this reason we derive the equation of motion for 
$\varphi(y)$ using the regularized action, eq. (\ref{regl}), from which we get:
\be
\chi_\lambda\left(
 \chi_{\lambda}\partial_{y}^{2}\varphi +
 2~\chi_{\lambda}'\partial_{y}\varphi +
 \chi_{\lambda}''\varphi +
 \chi_{\lambda} m^{2} \varphi\right) = 0~~~,
\ee
where we identified $\partial_{\mu}\partial^{\mu}\varphi$ with $m^{2}\varphi$.
The term in brackets should vanish everywhere, since it is continuous
and we can choose $\chi_\lambda(y)$ different from zero everywhere
except at one point between 0 and $2\pi R$.
If we finally take the limit $\lambda\to 0$
we obtain the equation of motion for the discontinuous fields:
\be
\label{discf}
 \chi \partial_{y}^{2}\varphi 
 - 4 \delta_{\pi R} \partial_{y}\varphi 
 - 2 \delta_{\pi R}' \varphi +
 \chi~ m^{2} \varphi = 0~~~.
\ee
Away from the point $y=\pi R$ this equation reduces to the equation of 
motion for continuous fields: terms with delta functions disappear and 
we can divide by $\chi(y)$. We obtain:
\be
 \partial_{y}^{2}\varphi +
 m^{2} \varphi = 0~~~.
\ee
Moreover, by integrating eq. (\ref{discf}) and its primitive 
around $y = \pi R$, we find:
\be
 \begin{array}{l}
  \varphi (\pi R + \xi ) = - \varphi (\pi R - \xi )\\
  \varphi '(\pi R + \xi ) = - \varphi '(\pi R - \xi )~~~,
 \end{array}
\ee
which are just the expected jumps. 

There is another possibility to 
derive the correct equation of motion from a singular action,
beyond that of adopting a convenient regularization.
We illustrate this procedure in the case of one complex scalar field
$\varphi(y)$.
The basic idea is to use a set of field variables such that 
their infinitesimal variations, implied by the action principle, 
are continuous functions of $y$. The action principle
requires that the variation of the action $S$, assumed 
to be a smooth functional of $\varphi$ and $\partial\varphi$, vanishes 
for infinitesimal variations of the fields from the classical trajectory:
\be
\delta S=\int d^4x~ dy~ \frac{\delta\mathcal{L}}{\delta\varphi} 
~\delta\varphi=0~~~.
\ee
If the system is described by discontinuous fields, in general we cannot
demonstrate that $\delta\mathcal{L}/\delta\varphi$ vanishes at the 
singular points, since multiplication/division by discontinuous functions 
like $\delta\varphi$ is known to produce inequivalent equalities. 
An exception is the case of
fields whose generic variation $\delta\varphi$ is a continuous function,
despite the discontinuities of $\varphi$. In this case the
action principle leads directly to the usual equation of motion.

We consider as an example the case discussed at the beginning of section 4.
Real and imaginary components of $\varphi$ are respectively even and
odd functions of $y$ and we have boundary conditions specified by
the matrices $U_\gamma$ in eq. (\ref{rotation}). 
In particular, the discontinuities of $\varphi_i$ $(i=1,2)$ and its
$y$-derivative across 0 and $\pi R$ are given by:
\be
\left(
\begin{array}{c}
\varphi_{1(2)}\\
\partial_y\varphi_{1(2)}
\end{array}
\right)(\gamma^+)
-
\left(
\begin{array}{c}
\varphi_{1(2)}\\
\partial_y\varphi_{1(2)}
\end{array}
\right)(\gamma^-)
=
(-)2
\tan\frac{\delta_\gamma}{2}
\left(
\begin{array}{c}
\varphi_{2(1)}\\
\partial_y\varphi_{2(1)}
\end{array}
\right)(\gamma)~~~,
\label{disc1}
\ee
where $\gamma$ stands for $0$ or $\pi R$, 
$\gamma^{+(-)}$ denotes $0^{+(-)}$ or $\pi R^{+(-)}$ and
\be
\left(
\begin{array}{c}
\varphi_{1(2)}\\
\partial_y\varphi_{1(2)}
\end{array}
\right)(\gamma)
\equiv
\frac{1}{2}
\left[
\left(
\begin{array}{c}
\varphi_{1(2)}\\
\partial_y\varphi_{1(2)}
\end{array}
\right)(\gamma^+)
+
\left(
\begin{array}{c}
\varphi_{1(2)}\\
\partial_y\varphi_{1(2)}
\end{array}
\right)(\gamma^-)
\right]~~~.
\ee
>From this we see
that a generic variation of $\varphi_2$ is discontinuous. The jump
of $\delta\varphi_2$ across $0$ or $\pi R$ is proportional to the 
value of $\delta\varphi_1$ at that point, which in general is not zero.
However we can move to a new set of real fields $\theta$ and $\rho$:
\be
\varphi =\rho~ e^{i \theta}~~~,
\ee
whose discontinuities from eq. (\ref{disc1}) read:
\be
\left(
\begin{array}{c}
\rho\\
\partial_y\rho
\end{array}
\right)(\gamma^+)
=
\left(
\begin{array}{c}
\rho\\
\partial_y\rho
\end{array}
\right)(\gamma^-)~~~~~~~~~~
\left(
\begin{array}{c}
\theta\\
\partial_y\theta
\end{array}
\right)(\gamma^+)
-
\left(
\begin{array}{c}
\theta\\
\partial_y\theta
\end{array}
\right)(\gamma^-)
=
\left(
\begin{array}{c}
\delta_\gamma\\
0
\end{array}
\right)~~~.
\label{disc2}
\ee
The discontinuity of $\theta$
at each fixed point is a constant, independent from
the value of $\varphi$ at that point. As a consequence, the infinitesimal 
variation $\delta\theta$ relevant to the action principle is continuous 
everywhere, including the points $y=0$ and $y=\pi R$.
We can derive the action for $(\rho,\theta)$, by starting from the Lagrangian
expressed in terms of $\varphi^c\equiv\rho\exp[i(\theta+\alpha)]$, where the 
function $\alpha$ has been defined in eq. (\ref{alpha}):
\be
\mathcal{L}(\varphi^c,\partial\varphi^c) = 
-\partial_{M}\varphi^{c\dag} \partial^{M}\varphi^c~~~. 
\ee
In terms of $\rho$ and $\theta$ we have:
\be
\mathcal{L}(\rho,\partial\rho,\theta,\partial\theta) = 
-\partial_{M}\rho\ \partial^{M}\rho -
\rho^{2}\partial_{M}(\theta +\alpha)\ \partial^{M}(\theta +\alpha)~~~.
\ee
The Lagrangian now contains singular terms, localized at the fixed points.
The equations of motions, derived from the variational principle, read:
\be
\left\{\begin{array}{l}
\partial_{M}\partial^{M}\rho - 
\rho\ \partial_{M}(\theta +\alpha)\ \partial^{M}(\theta +\alpha)=0\\
\partial_{M} [\rho^{2}\partial^{M}(\theta +\alpha)]=0
\end{array}\right. ~~~.
\label{ch}
\ee
In the bulk $\alpha$ is constant and drops from the previous equations,
which then become identical to the equations for the continuous field
$\varphi^c$, in polar coordinates. 
Moreover, by integrating eq. (\ref{ch}) around the fixed points and 
by recalling the properties of the function $\alpha$,  we 
reproduce precisely the jumps of eq. (\ref{disc2}).
The same results can be obtained by introducing a regularization for
$\alpha$.

In the previous examples we have dealt with free theories. 
It is interesting to examine what happens when interactions
are turned on. This is the case of non-abelian gauge theories.
The presence of derivative interactions and discontinuous fields
allows in principle localized interaction terms, that would provide
a non-trivial extension of the framework considered up to now.
To investigate this point we start from the 5D SU(2) Yang-Mills theory
defined by $Z=diag(+1,+1,+1)$ and by the boundary conditions in eq. (\ref{bcg1}):
\be
U_\beta = diag(-1,-1,+1)~~~,~~~~~~~~~~U_0=U_\pi=diag(+1,+1,+1)~~~.
\label{bc61}
\ee 
No jumps are present and the overall lagrangian is given only by the
`bulk' term:
\be
\label{lagg}
\mathcal{L} = -\frac{1}{4} F^{a}_{MN} F^{a MN}~~~.
\ee
It is particularly convenient to discuss the physics in the unitary gauge,
where all the would-be Goldstone bosons, eaten up by the massive
Kaluza-Klein modes, vanish:
\be
\label{ugauge}
A_5^a(x,y)\equiv 0~~~~~~~(a=1,2,3)~~~.
\ee
In this gauge $F^{a}_{5\mu}\equiv\partial_y A^a_\mu$ and the lagrangian (\ref{lagg}) reads:
\be
\label{laggug}
\mathcal{L} = -\frac{1}{4} F^{a}_{\mu\nu} F^{a \mu\nu}
-\frac{1}{2}\partial_y A^a_\mu \partial_y A^{a \mu}~~~.
\ee
To discuss the case of discontinuous gauge vector bosons, such as those
associated to the boundary conditions:
\be
U_\beta = U_0=diag(+1,+1,+1)~~~,~~~~~~~~~~U_\pi=diag(-1,-1,+1)~~~,
\label{bc62}
\ee
we can simply perform the field redefinition:
\be
A^{\hat{a}}_M(x,y)~\to~ \chi_\lambda(y)~ A^{\hat{a}}_M(x,y)~~~~~~({\hat{a}}=1,2)~~~,
\label{red6}
\ee
where the function $\chi_\lambda(y)$ represents a regularized version of
$\epsilon(y/2+\pi R/2)$. Such redefinition maps the twisted, smooth fields
obeying (\ref{bc61}) into periodic variables, discontinuous at $y=\pi R$, 
as specified in (\ref{bc62}). Notice that this redefinition does not change 
the gauge condition (\ref{ugauge}). If we plug the transformation (\ref{red6}) into the 
lagrangian (\ref{laggug}), we obtain the lagrangian for the system characterized 
by boundary conditions (\ref{bc62}). We stress that, since the field redefinition
(\ref{red6}) is local, the physics remains the same: the two systems
are completely equivalent. The S-matrix elements computed with the two
lagrangians are identical and, of course, this equivalence includes
the non-trivial non-abelian interactions.
Our aim here is only to understand how the
physics, in particular the non-abelian interactions,
are described by the new, discontinuous variables.

{}From eq. (\ref{laggug}) we can already conclude that
no localized non-abelian interaction terms arise from the field redefinition (\ref{red6}), 
in the unitary gauge. Indeed, the only term containing a $y$ derivative is 
quadratic, and, after the substitution (\ref{red6}), we will obtain terms analogous
to those discussed in (\ref{regl}) for the case of a single scalar field.
We find:
\begin{eqnarray}
\label{last-lagrangian}
\mathcal{L}& =
& -\frac{1}{4} \tilde{F}^{3}_{\mu\nu} \tilde{F}^{3 \mu\nu}
-\frac{1}{2}\partial_y A^3_\mu \partial_y A^{3 \mu} +  
\nonumber \\[0.15cm] 
&+&\frac{1}{2} \chi_\lambda^{2} f^{3\hat{b}\hat{c}} A^{\hat{b}}_{\mu} A^{\hat{c}}_{\nu}
   \tilde{F}^{3 \mu\nu} -\frac{1}{4} \chi_\lambda^{4} f^{3\hat{b}\hat{c}} f^{3\hat{d}\hat{e}} 
   A^{\hat{b}}_{\mu} A^{\hat{c}}_{\nu} A^{\hat{d}\mu} A^{\hat{e}\nu} -  
\nonumber \\[0.15cm] 
&-&\frac{1}{4} \chi_\lambda^{2} F^{\hat{a}}_{\mu\nu} F^{\hat{a}\mu\nu} 
-\frac{1}{2}\chi_\lambda^{2} \partial_y A^{\hat{a}}_\mu \partial_y A^{{\hat{a}} \mu} - 
\nonumber \\[0.15cm] 
& -&\frac{1}{2} ~\chi_\lambda'^{2}~ 
   A^{\hat{a}}_{\mu} A^{\hat{a}\mu} 
- \chi_\lambda \chi_\lambda'~A^{\hat{a}\mu}~ \partial_{5} A^{\hat{a}}_{\mu}
~~~,
\end{eqnarray} 
where $\tilde{F}^{3}_{\mu\nu} = \partial_{\mu} A^{3}_{\nu} -\partial_{\nu}
A^{3}_{\mu}$, $f^{abc}$ are the structure constants of SU(2) and the indices
$\hat{a},\hat{b},...$ run over 1,2. In the limit $\lambda\to 0$, the non-abelian 
interactions are formally unchanged, whereas the last line represents a set of
localized quadratic terms. As discussed in the case of a real scalar field,
such terms guarantee, via the equations of motion, that the fields obey the new 
boundary conditions (\ref{bc62}).
In a general gauge, interactions term localized at $y=\pi R$ are present,
but they involve non-physical would-be Goldstone bosons.

To summarize,
when going from a smooth to a discontinuous description of the 
same physical system, singular terms are generated in the
lagrangian. In our free theory examples as well in the non-abelian case we have 
quadratic terms localized at the orbifold fixed point.
Despite their highly singular behaviour, these terms
are necessary for a consistent description of the system.
Indeed they encode the discontinuities of the adopted field variables,
which can be reproduced via the classical equation of motion,
after appropriate regularization or through a careful 
application of the standard variational principle.
Conversely, when localized
terms for bulk fields are present in the 5D lagrangian, as 
for many phenomenological models currently discussed in 
the literature, the field variables are affected by discontinuities.
These can be derived by analyzing the regularized equation of motion 
and can be crucial to discuss important physical
properties of the system, such as its mass spectrum.
In some case we can find a field redefinition that eliminate
the discontinuities and provide a smooth description of
the system.



\vspace{0.5cm}

\bc
{\bf 7. Conclusion}
\ec
Symmetries and their realizations are central to our  
understanding of particle interactions. The possibility of
extending the present framework into a unified and more 
fundamental picture relies mostly on a realistic and consistent 
description of the breaking of viable symmetry candidates 
such as supersymmetry or grand unified symmetry.
     
In this respect conventional 4D field theories often appear 
to be quite limited. For example, it is well-known that 
a satisfactory 4D grand unified theory requires a rather baroque
Higgs sector, to overcome the long standing problem of doublet-triplet
splitting and the more recent one raised by the experimental bounds 
on the proton lifetime \cite{realgut}. The study of field theories with extra, 
compactified dimensions proved to be extremely fruitful
for the new possibilities offered for symmetry breaking.
One of the most interesting mechanism for symmetry breaking,
having no counterpart in four dimension, is that related to 
coordinate dependent-compactification, originally proposed by
Scherk and Schwarz. The lagrangian is invariant under 
a certain group of transformations, which, however, are not 
preserved by the boundary conditions obeyed by the fields. The 
extension of the Scherk-Schwarz proposal to the case of orbifold
compactifications has revealed several subtleties. The twist
defining the boundary conditions should obey certain
consistency conditions and the field variables are allowed to
possess discontinuities at the orbifold fixed points, or, saying
it differently, bulk fields are allowed to have lagrangian terms
localized at the fixed points.

In this paper we have given a systematic discussion of bosonic fields 
on an extra dimension compactified on $S^1/Z_2$. 
We were motivated by the following questions.
What kind of boundary conditions can be assigned to the fields?
We have seen that the orbifold constructions allows to
introduce jumps for both fields and first derivatives,
characterized by certain unitary transformations.
How are the boundary conditions related to the mass spectrum?
We found that the mass spectrum depends on a combination
of twist and jumps.
Does a given physical system correspond to a unique choice
of boundary conditions or can different boundary conditions
give rise to the same physics?
We found that there are local field redefinitions that relate
different boundary conditions, turning a jump into a twist or 
vice-versa. Therefore, within the examples explored, an entire
class of boundary conditions corresponds to the same physical 
system. In particular we found that, within this class, it is always 
possible to move to smooth field variables, where all the
relevant information is encoded in the twist. When using this set of field
variables, the lagrangian has only a `bulk' term. As expected, localized and
highly singular terms are present and actually needed when a description in
terms of discontinuous variables is adopted. At least within the class
of models explored here, such a description is not compulsory
and the localization of the corresponding lagrangian terms has no 
intrinsic physical meaning. 

\vspace*{1.0cm}
{\bf Acknowledgements}
We would like to thank A. Torrielli for useful discussions and 
J. Bagger and F. Zwirner for discussions,
suggestions and for their participation in the early stage of 
this work. C.B. and F.F. are partially
supported by the European Programs HPRN-CT-2000-00148 and HPRN-CT-2000-00149.
\vfill

\newpage
%

%
\end{document}